\documentclass[12pt]{article}
\usepackage{graphicx}
\def\hybrid{\topmargin 0pt      \oddsidemargin 0pt
        \headheight 0pt \headsep 0pt
       \voffset-1cm
        \textwidth 6.25in       % A4 paper
       \textheight 9.5in       % A4 paper
        \marginparwidth 0.0in
        \parskip 5pt plus 1pt   \jot = 1.5ex}
\catcode`\@=11
\def\marginnote#1{}

\newcount\hour
\newcount\minute
\newtoks\amorpm
\hour=\time\divide\hour by60
\minute=\time{\multiply\hour by60 \global\advance\minute by-\hour}
\edef\standardtime{{\ifnum\hour<12 \global\amorpm={am}%
        \else\global\amorpm={pm}\advance\hour by-12 \fi
        \ifnum\hour=0 \hour=12 \fi
        \number\hour:\ifnum\minute<10 0\fi\number\minute\the\amorpm}}
\edef\militarytime{\number\hour:\ifnum\minute<10 0\fi\number\minute}

\def\draftlabel#1{{\@bsphack\if@filesw {\let\thepage\relax
   \xdef\@gtempa{\write\@auxout{\string
      \newlabel{#1}{{\@currentlabel}{\thepage}}}}}\@gtempa
   \if@nobreak \ifvmode\nobreak\fi\fi\fi\@esphack}
        \gdef\@eqnlabel{#1}}
\def\@eqnlabel{}
\def\@vacuum{}
\def\draftmarginnote#1{\marginpar{\raggedright\scriptsize\tt#1}}

\def\draftlabel#1{{\@bsphack\if@filesw {\let\thepage\relax
   \xdef\@gtempa{\write\@auxout{\string
      \newlabel{#1}{{\@currentlabel}{\thepage}}}}}\@gtempa
   \if@nobreak \ifvmode\nobreak\fi\fi\fi\@esphack}
        \gdef\@eqnlabel{#1}}
\def\@eqnlabel{}
\def\@vacuum{}
\def\draftmarginnote#1{\marginpar{\raggedright\scriptsize\tt#1}}

\def\draft{\oddsidemargin -.5truein
        \def\@oddfoot{\sl preliminary draft \hfil
        \rm\thepage\hfil\sl\today\quad\militarytime}
        \let\@evenfoot\@oddfoot \overfullrule 3pt
        \let\label=\draftlabel
        \let\marginnote=\draftmarginnote
   \def\@eqnnum{(\theequation)\rlap{\kern\marginparsep\tt\@eqnlabel}%
\global\let\@eqnlabel\@vacuum}  }

\def\numberbysection{\@addtoreset{equation}{section}
        \def\theequation{\thesection.\arabic{equation}}}

\def\underline#1{\relax\ifmmode\@@underline#1\else
        $\@@underline{\hbox{#1}}$\relax\fi}

\def\titlepage{\@restonecolfalse\if@twocolumn\@restonecoltrue\onecolumn
     \else \newpage \fi \thispagestyle{empty}\c@page\z@
        \def\thefootnote{\fnsymbol{footnote}} }

\def\endtitlepage{\if@restonecol\twocolumn \else  \fi
        \def\thefootnote{\arabic{footnote}}
        \setcounter{footnote}{0}}  %\c@footnote\z@ }
\relax

\hybrid

\newfont{\Bbb}{msbm10 scaled 1\@ptsize00}
\newfont{\Bbbb}{msbm7 scaled 1\@ptsize00}

\newcommand{\DDD}{\raise-1pt\hbox{$\mbox{\Bbbb D}$}}

\newcommand{\UUU}{\raise-1pt\hbox{$\mbox{\Bbbb U}$}}

\newcommand{\z}{\raise-1pt\hbox{$\mbox{\Bbbb Z}$}}

\def\beq{\begin{equation}}
\def\eeq{\end{equation}}
\def\p{\partial}

\begin{document}

\pagenumbering{arabic}

\begin{titlepage}

\title{Bethe ansatz and Hirota equation  \\
in integrable models}

\author{A. Zabrodin
\thanks{Institute of Biochemical Physics,
4 Kosygina, 119334, Moscow, Russia; ITEP, 25 B. Cheremushkinskaya,
117218, Moscow, Russia; National Research University Higher
School of Economics,
20 Myasnitskaya Ulitsa, Moscow 101000, Russia}}

\date{October 2012}
\maketitle

\vspace{-7cm} \centerline{ \hfill ITEP-TH-48/12} \vspace{7cm}

\begin{abstract}

In this short 
review the role of the Hirota equation and the tau-function
in the theory of classical and quantum integrable systems is outlined.

\end{abstract}

\end{titlepage}

\paragraph{1. Introduction.}

As is known, integrable models exist in two versions -- 
classical and quantum.
In the former one should solve equations of motion while in
the latter the primary concern is to diagonalize operators
(Hamiltonians or evolution operators).
Usually classical integrable models admit quantizaton
that preserves integrability and quantum models have 
a well-defined 
classical limit, in accordance with the quantum-mechanical
correspondence principle.

At the same time
it turns out that there are deeper links
between quantum and classical integrable models
which do not follow from the usual correspondence principle.
Namely, it appears that classical integrable equations
are built in the structure of quantum models as exact
relations even at
$\hbar \neq 0$. And vice versa, some specific
accessories of the quantum theory appear in solving
purely classical integrable problems.
There are several aspects of this surprising and not yet fully understood
phenomenon.
We discuss only one -- the identification of the quantum
transfer matrix
(the $T$-operator) with the classical $\tau$-function.

In the main text we do not give any references.
Some references with brief comments are collected in section 7.

\vspace{-5mm}

\paragraph{2. Quantum integrable systems: Bethe ansatz.}

It is noteworthy that the first non-trivial 
problem with many degrees of freedom solved exactly
was quantum rather than classical.
In 1931 H.Bethe managed to find exact wave functions of
the Heisenberg spin chain with the Hamiltonian
$$
H=J\sum_{n=1}^{L} \left (\sigma_{n}^{x}\sigma_{n+1}^{x}+
\sigma_{n}^{y}\sigma_{n+1}^{y}+
\sigma_{n}^{z}\sigma_{n+1}^{z}\right ).
$$
Here $\sigma^x , \sigma ^y, \sigma^z$ are the standard Pauli 
matrices.
The key point of the solution
was a special ansatz for the wave function which is now
called the Bethe ansatz. Later one or other version of
this method turned out to be applicable to many other
lattice or continuous quantum models
in $1+1$ dimensions. The typical form of the answer for
the spectrum of the Hamiltonian
is
$$
E=\sum_{k} \varepsilon (v_k).
$$
Here $\varepsilon (v)$ is a known function
(for the Heisenberg chain it is $\varepsilon (v)=
-8J/(v^2 +1)$) and the numbers $v_k$ which
are rapidities of quasiparticles are found from the
system of algebraic equations
(the Bethe equations)
\beq\label{beq1}
\left (\frac{v_k -i}{v_k +i}\right )^L =
-\prod _{l}\frac{v_k -v_l -2i}{v_k -v_l +2i}\,.
\eeq
In physically interesting cases, when the ground state
in the thermodynamic limit is built by filling a false
vacuum and the number of quasiparticles
tends to infinity together with the lattice length $L$,
the system of Bethe equations becomes an integral
equation for the density of continuously distributed
rapidities which can be solved explicitly.

In the case of generalized spin chains with ``spin variables''
belonging to representations
of the group $SU(N)$, diagonalization of Hamiltonian
is performed by means of consecutive application
of the Bethe ansatz
$N-1$ times (this method is usually referred to as nested Bethe ansatz).
Correspondingly, there are several sorts of quasiparticles
with rapidities $v_{k}^{(t)}$, $t=1, 2, \ldots , N-1$,
and the Bethe equations read
\beq\label{beq2}
\prod_l \,
\frac{v_{k}^{(t)}-v_{l}^{(t-1)}+i}{v_{k}^{(t)}-v_{l}^{(t-1)}-i}\,\,\,
\frac{v_{k}^{(t)}-v_{l}^{(t)}-2i}{v_{k}^{(t)}-v_{k}^{(t)}+2i}\,\,\,
\frac{v_{k}^{(t)}-v_{l}^{(t+1)}-i}{v_{k}^{(t)}-v_{l}^{(t+1)}+i}\,
=\, -1.
\eeq

For many years a number of different versions
and generalizations of the Bethe method were suggested.
However, its main secret seems to be still unrevealed.

\vspace{-5mm}

\paragraph{3. Integrable models of classical theory:
Hirota equation.}

It is a question about non-linear partial differential equations.
Today we know many cases when they appear to be integrable.
Among them are well-known examples of integrable models
of classical field theory in $1+1$ dimensions:
sin-Gordon, the principal chiral field, sigma-models.
There are also non-relativistic models such as
the Korteweg - de Vries equation, the non-linear Schrodinger
equation, the Kadomtsev-Petviashvili equation, the Toda chain
and many other. They are often called soliton equations
because they usually have exact solutions of that
type (localized moving excitations that preserve their shape
in the evolution). The integrability, i.e. the existence
of infinite number of conserved quantities in involution,
means that with each such equation one can associate
an infinite
{\it hierarchy} of compatible equations since
each integral of motion generates its own evolution
in time.

The central object of the theory is a function
on an infinite dimensional parameter space
which is called $\tau$-function.
In a nutshell, it provides a common solution to the whole
hierarchy of integrable equations generated by the infinite
family of Hamiltonians in involution.
As such, it depends on infinite number of variables
${\sf t}=\{t_0,
t_1, t_2, t_3, \ldots \}$ and satisfies infinite number of
equations which can be encoded in just one functional relation
(the Hirota equation)
\beq\label{Hir1}
\begin{array}{l}
(z_2 \! -\! z_3)\, \tau ({\sf t} +[z_1])
\tau ({\sf t} +[z_2]+[z_3])\\ \\
\hspace{10mm}+\,\,
(z_3 \! -\!z_1)\, \tau ({\sf t} +[z_2])
\tau ({\sf t} +[z_1]+[z_3])\\ \\
\hspace{20mm}+\,\,
(z_1 \! -\!z_2)\, \tau ({\sf t} +[z_3])
\tau ({\sf t} +[z_1]+[z_2]) \,\, =\,\, 0.
\end{array}
\eeq
where $z_1, z_2 , z_3$ are arbitrary
parameters, and we use the short-hand notation
${\sf t} +[z ] \equiv \{t_0+1,
t_1 +z,
t_2 +\frac{1}{2}z^2, t_3
+\frac{1}{3}z^3 , \ldots \}$. Expansion in powers of
$z_i$ yields differential equations which constitute the
hierarchy.
(To be precise, (\ref{Hir1})
corresponds to the modified Kadomtsev-Petviashvili
hierarchy; in other cases the functional relations
have a similar bilinear form.)
Along with $t_i$ one may also use
the variables $u_{z}$ which are numbered by
continuous ``label'' $z$ and which are connected with
the $t_i$'s by the relations
\beq\label{zamena}
t_0=u_0,
\quad \quad
t_k = \frac{1}{k}\sum_{z} u_{z}z^{k},
\quad \quad
k\geq 1
\eeq
(let us assume that the sum is finite).
In the variables $u_{z}$ equation (\ref{Hir1})
becomes a difference equation on a 3D lattice
for any triple $u_{z_i}=u_i$
$(i=1,2,3)$:
\beq\label{Hir2}
\begin{array}{l}
(z_2 \! -\! z_3)\, \tau (u_1, u_2 + 1, u_3 + 1)
\tau (u_1 + 1, u_2, u_3)\\ \\
\hspace{10mm}+\,\,
(z_3 \! -\!z_1)\, \tau (u_1  + 1, u_2, u_3 + 1)
\tau (u_1, u_2 + 1, u_3)\\ \\
\hspace{20mm}+\,\,
(z_1 \! -\!z_2)\, \tau (u_1  + 1,u_2 + 1,  u_3)
\tau (u_1, u_2,  u_3 + 1) \,\, =\,\, 0.
\end{array}
\eeq
The parameters $z_i$ play the role of the lattice spacings.
Similar equations can be written
for four and more variables
$u_i$ but all of them appear to be algebraic consequences
of equations (\ref{Hir2}) written for any triple of variables.

The sets of variables
$t_i$ and $u_{z}$ provide complimentary descriptions
of one and the same integrable system and transition
from one to another
is in some sense similar to the Fourier transform.
Let us also note that the Hirota equation written
in the forms (\ref{Hir1})
or (\ref{Hir2}) reflects a deep interrelation between
continuous and discrete or difference soliton equations:
they belong to one and the same hierarchy and turn one
into another at the change of the independent variables
(\ref{zamena}) made simultaneously in the whole hierarchy.

``In nature'' $\tau$-functions or their logarithms
appear as partition functions,
different kinds of correlators and their
generating functions, and effective actions as functions
of coupling constants.

In fact the whole variety of integrable non-linear partial differential
equations can be encoded in {\it one} universal
difference equation (\ref{Hir2}) for the
$\tau$-function. As well as the equivalent equation
(\ref{Hir1}), it is called the Hirota equation.
It plays a truly vital role in the theory of classical
(and also quantum, as we shall see soon)
integrable systems. All known integrable equations
can be obtained from it by various simple
but maybe technically sophisticated manipulations
such as continuous limit (expansion in powers of
$z_i$ and transition to the variables $t_k$),
imposing of reductions, choosing dependent and
independent variables and the like.

As any fundamental thing,
the Hirota equation and closely related equation called
the $Y$-system often appears in different unexpected contexts.
Recently it was used for finding the spectrum of anomalous
dimensions of composite operators in
$N=4$ su\-per\-sym\-met\-ric 4D Yang-Mills theory in the
planar limit.

\vspace{-5mm}

\paragraph{4. Towards a synthesis
of classical and quantum integrability.}

By quantum integrable system we mean, in this section,
a non-homogeneous spin chain whose local observables
(``spins'' on the sites) are operators acting in
finite dimensional representations
of the $su(N)$ algebra or its $q$-deformation.
This example is rather representative because
many other exactly solvable quantum models
can be treated, at least formally, as its limiting cases.

Since the system has many
or even infinitely many commuting integrals of motion $H_{\{J\}}$,
where $\{J\}=\{J_1, J_2, \ldots \}$ is some multi-index,
it is natural to
diagonalize them simultaneously rather than one particular
Hamiltonian from this family.
Even better, one can combine them in a generating function
and diagonalize this operator function.
Schematically, it looks like
\beq\label{Top}
T(t_0 , t_1 , t_2, \ldots )
= \sum_{\{J\}} \, a_{\{J\}}\Bigl (\prod_j t_{j}^{J_j}\Bigr )\,
H_{\{J\}},
\eeq
where $a_{\{J\}}$ are properly chosen coefficients.
It depends on an infinite number of auxiliary variables $t_i$.
Such generating function is called the master
$T$-operator (or simply $T$-operator). It has the meaning of
an evolution operator in a time determined by the parameters $t_i$.
The $T$-operator
is a much more meaningful and informative
object than any particular Hamiltonian
because an important dynamical information is encoded in its
analytic properties in the variables $t_i$ (which are
in general complex numbers).

The commuting family of operators constructed in such a way
contains, along with the full set of commuting Hamiltonians
of the spin chain, the transfer matrices of associated
2D lattice models of statistical mechanics (vertex models)
as well as all Baxter's $Q$-operators.

More often than not an explicit expression for the
$T$-operator through original dy\-na\-mi\-cal variables
is not available but this is not an obstacle for
the derivation of general functional relations for it.
With taking into account the analytic properties,
they allow one to solve the spectral problem for the
$T$-operator and thus for the original Hamiltonian.
These functional relations have a long history and
are known in different forms. In a sense, they are
at the top of
the theory of quantum integrable systems.
Remarkably, at this place the quantum theory becomes
very close to the classical one.

The key fact that provides a hint for the anticipated
synthesis of classical and quantum integrability
is that the most general and universal form of the
functional relations for the $T$-operator
$T(t_0, t_1, \ldots )$ (\ref{Top}) is nothing else
than the classical Hirota equation (\ref{Hir1})
in the variables $t_i$. Let us remark that since
the $T$-operators commute for all values of $t_i$,
the ordering ambiguity, usually accompanying
quantization, does not arise here, and any eigenvalue
of the $T$-operator satisfies the Hirota equation.
More precisely, the $T$-operator satisfies 
the general bilinear relation for the 
$\tau$-function of the modified Kadomtsev-Petviashvili (mKP)
hierarchy from which the equations of the Hirota type follow.
In other words, the $T$-operator
(or rather any of its eigenvalues) and the classical
$\tau$-function can be identified:
$$T(t_0, t_1, t_2, \ldots )=\tau (t_0 , t_1 , t_2 , \ldots ).$$
To make the statement more 
precise, we need some details about the $T$-operator.
Among the variables $t_i$, the first one, $t_0$,
is distinguished. In order to stress this,
we denote it by $u$: $u=t_0$.
It is called the quantum spectral parameter.
Let us consider the
$T$-operator as a function of $u$, depending on
all other ${\bf t}=\{t_1, t_2, \ldots \}$ as on parameters:
$T=T(u, {\bf t})$
(for definiteness, one can keep in mind any of its eigenvalues).
Let $T_{\lambda}(u)$ be the set 
of commuting quantum transfer matrices. They are functions of $u$ 
and depend on a Young diagram $\lambda$ as on discrete 
(multi-component) parameter.
The $T$-operator can be represented as the following generating
series which is the precise version of the 
schematic formula (\ref{Top}):
\beq\label{T-operator}
T(u, {\bf t})=\sum_{\lambda}s_{\lambda}({\bf t})
T_{\lambda}(u).
\eeq
The sum is taken over all Young diagrams including the 
empty one and $s_{\lambda}({\bf t})$ are the Schur functions
(which are usually regarded as symmetric functions of 
the variables $x_i$ such that $t_k =\frac{1}{k}\sum_i x_i^k$).
The known functional relations for $T_{\lambda}(u)$ imply that
this object satisfies the classical mKP hierarchy and, in particular,
the Hirota equation.
The Planck constant $\hbar$ of the quantum problem
is not present in the Hirota equation but appears
in its solutions.

In the context of the classical Hirota equation,
the nested Bethe ansatz method
translates to a chain of B\"acklund transformations which
sequentially ``undress'' the original solution to a trivial one.

All this can be almost literally extended to
quantum integrable models with supersymmetry.
The Hirota equation for the
$T$-operator remains the same, the only change is
the class of relevant solutions.

\vspace{-5mm}

\paragraph{5. Connection with many-body
problems of the Calogero-Moser type.}

Another side of the classical-quantum correspondence is
the link to classical
integrable many-body problems of the Calogero-Moser type
which enter into the game through dynamics of zeros
of the $T$-operator.
This link follows from the identification
of the $T$-operator
with the classical $\tau$-function and from the
analytic properties of the former.

To give a more precise statement, we need to know 
the analytic properties of the $T$-operator
$T(u, {\bf t})$, $u=t_0$, in the variable $u$ (the quantum
spectral parameter).
This is one of the most important characteristics of the quantum model.
In the simplest case, which includes finite spin chains,
the $T$-operator should be a polynomial of
$u$ of degree equal to the length of the chain:
\beq\label{zeros}
T(u, {\bf t})=C\prod_{j=1}^{L}(u-u_j({\bf t})).
\eeq
Then from the identification of $T(u, {\bf t})$ with the $\tau$-function
it follows that the dynamics of the zeros
$u_j({\bf t})$ in the times $t_i$
is given by the relativistic generalization of the
Calogero-Moser model
(it is usually called the Ruij\-se\-na\-ars-\-Schnei\-der system).
For example the equations of motion in $t_1$ are
\beq\label{ruij}
\ddot u_i =\sum_{k\neq i}
\frac{2\dot u_i \dot u_k}{(u_i \! -\! u_k)^2 -1}\,, \quad \quad
\dot u_i = \p_{t_1}u_i \,.
\eeq
This system is integrable and has the required number of
independent conserved quantities ${\cal H}_j$ in involution.

Finally, let us stress the specific way of posing the classical
mechanical problem for the Ruijsenaars-Schneider
system that corresponds to diagonalization
of quantum spin chain Hamiltonians or
transfer matrices of the vertex model.
The standard mechanical problem is: given initial coordinates
and velocities of the particles $u_j(0)$, $\dot u_j(0)$, find
$u_j(t)$. In distinction to this,
in order to find
eigenvalues of the $T$-operator one should pose the problem
in the following non-standard way: given initial coordinates
$u_j=u_j(0)$ and values of all higher integrals of motion ${\cal H}_j$,
find initial velocities $\dot u_j(0)$.
The solution of such a problem in general
is not unique: different possible solutions
correspond to different eigenstates of the quantum
Hamiltonians.

We also note that the Bethe equations
(\ref{beq2}) can be understood as
the Ruij\-se\-na\-ars-\-Schnei\-der system
in the discrete time $t$.

\vspace{-5mm}

\paragraph{6. Concluding remarks.}

We see that the most universal relations
for classical and quantum integrable systems actually
coincide and are given by the Hirota equation.
Summing up, we suggest the following extension
of the quantum-mechanical cor\-res\-pon\-dence principle:
with any quantum integrable system one can associate
a classical in\-teg\-rab\-le dynamics in the space of
its (commuting) integrals of motion.
This classical dynamics contains complete information
about spectral properties of the quantum system.

At this point we finish the story about
interrelations between classical and quantum integrability
told ``at verbal level''. In our opinion, the very fact that
the top points of the classical and quantum theories
of integrable systems actually coincide takes place for
profound reasons and
requires a deeper understanding.

\vspace{-5mm}

\paragraph{7. Some references and comments.}

The fundamental role of the $\tau$-function
in the theory of soliton equations
was revealed in
the works of the Kyoto school
(see e.g. \cite{DJKM83} and references therein).
The discrete Hirota equation first appeared in \cite{Hirota81},
its meaning for integrable hierarchies was clarified in
\cite{Miwa82}. The Bethe ansatz is reviewed in
\cite{KBIbook,Faddeev}. The functional relations
for quantum transfer matrices in the form of determinant formulas
were found in \cite{Chered,BR90}, for the supersymmetric 
extension see \cite{Tsuboi}.
A similarity between quantum transfer
matrices and classical $\tau$-functions was first pointed out in
\cite{KLWZ} (see also \cite{Z98}), where the discrete classical
Hirota dynamics in the space of commuting quantum integrals of motion
was introduced. An extension of this approach to supersymmetric
models was suggested in \cite{KSZ08}. For the master $T$-operator
see \cite{KLT10,AKLTZ11}. The dynamics of poles of solutions to 
classical integrable equations (zeros of the $\tau$-function) 
in connection with solvable many-body problems was studied in 
\cite{AMM,Krichever,Shiota}. The coincidence of the equations of motion
for the discrete time analog of the Ruijsenaars-Schneider system 
and the Bethe equations was noticed in \cite{NRK}, a connection
with motion of zeros of $\tau$-function was pointed out in
\cite{KLWZ}. 
For various aspects
and applications of the Hirota
equation (the $T$-system) and the associated $Y$-system
see review \cite{KNS11} and references therein.
Another context where $\tau$-functions
of classical integrable hierarchies enter the theory of
quantum integrable
models and associated 2D lattice models of statistical mechanics
is calculation of scalar products and partition functions
with domain wall boundary conditions \cite{dw}.

\vspace{-5mm}

\paragraph{Acknowledgements.}
The author thanks A.Alexandrov, A.Gorsky, V.Kazakov, I.Kri\-che\-ver,
S.Leu\-rent, A.Orlov, T.Ta\-ke\-be, Z.Tsuboi
for discussions and to A.Morozov for reading the manuscript.
This work was supported in part by RFBR grant
12-01-00525, by joint RFBR grants 12-02-91052-CNRS,
12-02-92108-JSPS, by grant NSh-3349.2012.2 for support of 
leading scientific schools and
by Ministry of Science and Education of Russian Federation
under contract 8206.

\end{document}